\documentclass{aa}  

\usepackage{color}
\usepackage{graphicx}
\usepackage{ragged2e}
\usepackage[varg]{txfonts}
\usepackage{multirow}
\usepackage{graphicx}

\begin{document}

   \title{The evolution of Balmer jump selected galaxies in the ALHAMBRA survey.
            \thanks{Based on data obtained at the Calar Alto Observatory}}

    \titlerunning{The evolution of Balmer jump selected galaxies in the ALHAMBRA survey.}

   \author{P. Troncoso Iribarren\inst{1,2},
	  L.~Infante\inst{1,2},
      N.~Padilla\inst{1,2},
	  I.~Lacerna\inst{1,2,5},
      S.~Garcia\inst{1,2},
  	  A.~Orsi\inst{3},
	  A.~Mu\~noz Arancibia \inst{1,\ref{uvp}},
	  J.~Moustakas\inst{4},
	  D.~Crist\'obal-Hornillos \inst{\ref{cefca}},
	  M.~Moles\inst{\ref{cefca},\ref{a2}},      
	  A.~Fern\'andez-Soto\inst{\ref{a5},\ref{a12}},
      V.~J.~Mart\'inez\inst{\ref{a5},\ref{a3},\ref{a4}},
	  M.~Cervi\~no\inst{\ref{a2},\ref{a10},\ref{a11}}	
      E.J.~Alfaro\inst{6},
	  B.~Ascaso\inst{\ref{a2},\ref{b3}},
	  P.~Arnalte-Mur\inst{\ref{a3}},
      L.~Nieves-Seoane\inst{\ref{a12},\ref{a3}}
   \and
	  N.~Ben\'itez\inst{\ref{a2}}
}
  \offprints{P. Troncoso Iribarren, ptroncos@astro.puc.cl}
   \institute{Instituto de Astrof\'isica, Universidad Cat\'olica de Chile, 
	Av. Vicuna Mackenna 4860, 782-0436 Macul, Santiago, Chile
	\and
	Centro de Astro-Ingenier\'ia, Universidad Cat\'olica de Chile,
     Av. Vicuna Mackenna 4860, 782-0436 Macul, Santiago, Chile.
   \and
	      Centro de Estudios de F\'isica del Cosmos de Arag\'on (CEFCA),
	      Plaza de San Juan 1, Teruel, 44001, Spain\label{cefca}.
	\and
	      Department of Physics and Astronomy, 
	      Siena College, 515 Loudon Road, Loudonville, NY 12211, USA.
	\and
		Max Planck Institute for Astronomy, K{\"o}nigstuhl 17, 
	    69117, Heidelberg, Germany.
	\and
		  Instituto de Astrof\'sica de Andaluc\'ia (IAA-CSIC),
          Glorieta de la Astronom\'ia, 18008-Granada, Spain\label{a2}.
	\and
		Instituto de Astrof\'isica de Canarias, 
        V\'ia L\'actea s/n, 38200, La Laguna, Tenerife, Spain\label{a10}.
    \and
		Departamento de Astrof\'isica, Facultad de F\'isica,
        Universidad de La Laguna, 38206 La Laguna, Spain\label{a11}.
    \and
		Unidad Asociada Observatorio Astron\'omico (IFCA-UV), 
         E-46980, Paterna, Spain\label{a5}.
    \and
		Instituto de F\'isica de Cantabria (CSIC-UC), 
        E-39005 Santander, Spain\label{a12}.
    \and    
        Observatori Astron\`omic, Universitat de Val\`encia, 
        C/ Catedr\`atic Jos\'e Beltr\'an 2, E-46980, Paterna, Spain\label{a3}.
	\and
	Departament d'Astronomia i Astrof\'isica, 
        Universitat de Val\`encia, E-46100, Burjassot, Spain\label{a4}.
                   \and
	GEPI, Observatoire de Paris, CNRS, Universit\'e Paris Diderot, 61, 
         Avenue de l’Observatoire 75014, Paris, France\label{b3}.
      \and  
      Instituto de F\'isica y Astronom\'ia, Universidad de Valpara\'iso, Avda. Gran Bretana 1111,
      Valpara\'iso, Chile \label{uvp}.
}

    \authorrunning{P. Troncoso Iribarren et al. }
	      
   \date{Received October, 2015; accepted }

% \abstract{}{}{}{}{} 
% 5 {} token are mandatory
   \abstract
  {Samples of star-forming galaxies at different redshifts have been traditionally selected via color techniques. The ALHAMBRA survey was designed to perform a uniform cosmic tomography of the Universe, and we will exploit it here to
    trace the evolution of these galaxies.}
   {Our objective is to use the homogeneous optical coverage of the ALHAMBRA filter system 
   to select samples of star-forming galaxies at different 
   epochs of the Universe and study their properties.}
   { We present a new color-selection technique,
   based on the Bruzual $\&$ Charlot 2003 models convolved with the ALHAMBRA bands,
   and the redshifted position of the Balmer jump 
   to select star-forming galaxies in the redshift range $0.5 < z < 1.5$.
   These galaxies are dubbed Balmer jump Galaxies ($BJGs$).
   We apply the {\it iSEDfit} Bayesian approach to fit each detailed SED 
   and determine star-formation rate (SFR), stellar mass, age and absolute magnitudes. 
   The mass of the haloes where these samples reside are found via a clustering analysis.}
  { Five volume-limited $BJG$ sub-samples with different mean redshifts are found to
  reside in haloes of median masses $\sim 10^{12.5 \pm 0.2}M_{\sun}$ slightly increasing toward $z=0.5$.
  This increment is similar to numerical simulations results which suggests that 
  we are tracing the evolution of an evolving population of haloes as they grow to reach a mass of $\sim 10^{12.7 \pm 0.1}$ at $z=0.5$.
  The likely progenitors of our samples at $z\sim$3 are Lyman Break Galaxies, which 
  at $z\sim$2 would evolve into star-forming $BzK$ galaxies,
  and their descendants in the local Universe are elliptical galaxies.
  Hence, this allows us to follow the putative evolution of the SFR, stellar mass and age 
  of these galaxies.}
  {From $z\sim 1.0$ to $z\sim 0.5$, the stellar mass of the volume limited $BJG$ samples
  nearly does not change with redshift, 
  suggesting that major mergers play a minor role on the evolution of these galaxies. 
  The SFR evolution accounts for the small variations of stellar mass, 
  suggesting that star formation and possible minor mergers 
  are the main channels of mass assembly. 
   }

   \maketitle

%
%________________________________________________________________

\section{Introduction}
%__________________________________________________________________
Most of the recent observational efforts to understand galaxy evolution
have been focused on determining the history of cosmic star formation, 
gas density evolution, metallicity evolution
and mass growth of the Universe \citep{daddi04,mannucci10,madau14,tomczak14,bouwens15}.
These multiwavelength observational constraints 
have been usually summarized as galaxy scaling relations 
that might change or not with redshift \citep{mannucci10,elbaz11,bouwens14,troncoso14},
in high or low density environments, in extreme physical conditions (starburst, AGN galaxies),
and in spatially resolved data due to internal variations of the galaxy properties \citep{sanchez13}.
In parallel, theoretical works and simulations
have tried to explain the physical mechanisms that reproduce the measured global properties
\citep{daddi10,dave11,lilly13,lagos14,padilla14}.
Despite these efforts, % the cosmic variance, 
the completeness and cleanness of the sample
are still challenging issues 
that depend on the sample selection-method, instruments limit, and telescope time.
The aforementioned issues make the comparison 
between observational and theoretical works even more difficult.
For example, \cite{campbell14} compared the stellar mass of GALFORM galaxies
predicted by the model with the ones obtained via the fit of their predicted broad-band colors.
They find that, for an individual galaxy, both quantities differ, 
hence the clustering of mass-selected samples can be affected by systematic biases.
Therefore, mass-selected samples might provide erroneous conclusions regarding their progenitors and descendants.
Besides, the evolution of scaling relations is constrained with observations 
of galaxy samples, selected with luminosity or stellar-mass thresholds,
located at different redshifts,
which does not necessarily constitute causally connected populations
(i.e. do not follow a progenitor-to-descendant relationship).
Clustering selected samples overcome this issue because,
in a hierarchical clustering scenario,
a correlation analysis allows us to estimate the bias 
and hence statistically determine the progenitors and descendants of galaxy samples.
The bias parameter measures the clustering difference between the galaxy spatial distribution 
and underlying dark-matter distribution.
Thus, it relates the typical mass of haloes hosting the galaxies \citep{sheth_mo_tormen01}.
Hence, measuring it in galaxy samples at different redshifts,
determines whether we are following the evolution of baryonic processes occurring on haloes of similar masses or not.
This fact is of extreme importance because once it is determined
we can use the multiwavelength data to study the evolution 
of the baryonic processes at certain halo mass,
establishing a direct link between observations and galaxy formation models.
\cite{padilla10} selected early-type galaxies according to their clustering and 
luminosity function in the MUSYC survey.
So far, no study that selects star-forming galaxies according their clustering and luminosity function has been reported.

Star-forming galaxies are of particular interest,
because they allow us to study the mechanisms that switch on/off the star formation 
and its evolution with redshift.
Considering the lack of wide spectroscopic surveys,
in the sense of wavelength coverage and surveyed area,
the majority of the star-forming galaxy samples 
have been chosen using two-color selection techniques.
The so-called ``dropout'' technique is based on recording 
the difference between two distinct parts of the spectrum,
which generate a break on it (e.g. the 912\AA~Lyman break, the 3646\AA~Balmer jump).
This difference is strong enough that it has been measured in broad bands, 
selecting star-forming galaxies at early periods of the Universe ($z>1.4$), 
%broadening galaxy taxonomy 
such as the BzK, BX, BM, DOGs and LBGs \citep{steidel96,daddi04,steidel04,dey08,infante15}.
Several authors have measured the bias \citep{gawiser07,blanc08,guaita10}  determining the mass of the halo where each galaxy sample resides
and connecting the progenitors and descendants of these galaxy samples.
Other works selected the samples trusting
fully on their photometric-redshift and
the physical properties determined via SED-fitting \citep{tomczak14},
far-IR luminosity \citep{rodighiero11}
or Bayesian approaches such as {\tt iSEDfit} \citep{moustakas13}.
Recently, \cite{viironen15} implemented a method in the ALHAMBRA survey
to select galaxy samples using
the probability distribution of the photometric redshift (zPDF).
The quality of the detailed SED distribution,
provided by the medium bands of the ALHAMBRA survey,
allowed them to perform an accurate statistical analysis.
Indeed, they include our lack of knowledge on the precise galaxy redshift and select the sample according to {\it certain probability threshold} defined by the authors.
Therefore, by definition this method selects a clean but not a {\it complete} sample.

In this work, we aim to develop a technique,
{\it base purely on photometric data},
to select star-forming galaxies.
This kind of selection allows us to directly compare with
the previously mentioned works that also use a dropout technique
to select their BzK, LBG, etc. samples.
We use the uniform separation between two contiguous medium-bands
of the ALHAMBRA survey to register the redshifted position of the 
3646\AA~ Balmer jump within the optical domain, 
allowing us to select galaxy samples in the redshift range $0.5 < z < 2$. 

We base our two-color selection technique on the \cite{bc03} models
and apply it to the GOLD ALHAMBRA catalogs \citep{molino13}.
In the following, the galaxies selected by this method are dubbed 
Balmer jump Galaxies (BJGs) and their physical properties are investigated.
Based on a clustering study, 
we find the progenitors and descendants of galaxies in these samples,
allowing us to study the evolution of the SED fit derived properties 
as a function of redshift on haloes of certain mass.
This paper is organized as follows:
in section \S\ref{s_alh}, we summarize the ALHAMBRA observations and
introduce the nomenclature of the ALHAMBRA filter system used throughout the paper.
In section \S\ref{s_sample}, the selection method is described and attested with the \cite{bc03} models.
The $BJG$ samples are defined here.
In section \S\ref{s_pp}, each galaxy SED is modeled using {\tt iSEDfit} \citep{moustakas13}
and the physical properties of each sample are characterized as a whole.
In section \S\ref{s_clustering}, the clustering properties are calculated.
In section \S\ref{s_discussion}, the main results are discussed.
Finally, in section \S\ref{s_conclusions} our conclusions are exposed.
Throughout the paper, we use a standard flat cosmology
with $H_0=100 h \,km \, s^{-1} \, Mpc^{-1}$, $\Omega_m(z=0)=0.3$, $\Omega_{\Lambda}(z=0)=0.7$,
$\sigma_8=0.824 \pm 0.029$, and the magnitudes are expressed in the AB system.

\begin{table}
\begin{center}
\label{table_fname}
\caption{Name and effective wavelength of each ALHAMBRA filter.}
\begin{tabular}{lcccccc}
\hline
\hline
Name &$\lambda_{eff}$  &Name &$\lambda_{eff}$ &Name &$\lambda_{eff}$ \\
     & [nm]	&&[nm]		&	&[nm]\\
%     \\
\hline
\noalign{\smallskip}
$U_1$	 & 365.5 	&$U_2$    & 396.5	&$B_3$	  & 427.5\\
\noalign{\smallskip}
$B_4$	 & 458.5 	&$B_5$	  & 489.5	&$B_6$	  & 520.5\\	
\noalign{\smallskip}
$B_7$	 & 551.5	&$R_8$	  & 582.5	&$R_9$	  & 613.5\\
\noalign{\smallskip}
$R_{10}$ & 644.5	&$R_{11}$ & 675.5 	&$R_{12}$ & 706.5\\
\noalign{\smallskip}
$I_{13}$ & 737.5	&$I_{14}$ & 768.5	&$I_{15}$ & 799.5\\
\noalign{\smallskip}
$I_{16}$ & 830.5	&$z_{17}$ & 861.5	&$z_{18}$ & 892.5\\
\noalign{\smallskip}
$z_{19}$ & 923.5	&$z_{20}$ & 954.5 	& 	  &\\
\hline 
\end{tabular}
\end{center}
{\bf Notes:} Cols.1, 3, and 5 indicate the adopted filter name 
that will be used throughout the paper ;
Cols.2, 4, and 6 show the effective wavelength of each ALHAMBRA band.

\end{table}
%__________________________________________________________________

\section{Data: the ALHAMBRA survey
%\footnote[1]{\tt http://www.alhambrasurvey.com}
}
\label{s_alh}

%__________________________________________________________________

The Advanced Large Homogeneous Area Medium-Band Redshift Astronomical (ALHAMBRA\footnote[1]{\tt http://www.alhambrasurvey.com})
survey provides a kind of {\it cosmic tomography} for the evolution of 
the contents of the Universe over most of the cosmic history. 
\cite{benitez09} have especially designed a new optical photometric system for the ALHAMBRA survey,
which maximizes the number of objects with an accurate classification by the
Spectral Energy Distribution (SED) and photometric redshift.
It employs 20 contiguous, equal-width $\sim$ 310 \AA, 
medium-band covering the wavelength range from 3500 \AA$\,$ to 9700 \AA,
plus the standard $JHK$ near-infrared bands. 
\cite{moles08}, and \cite{apariciovillegas} presented an extensive description of the survey
and filter transmission curves.
The observations were taken in the Calar Alto Observatory (CAHA, Spain)
with the 3.5-m telescope using 
the two wide-field imagers in the optical (LAICA) and NIR (Omega-2000).
The total surveyed area is 2.8deg$^2$ distributed in eight fields that overlap areas 
of other surveys such as SDSS, COSMOS, DEEP-2, and HDF-N.
The typical seeing of the optical images is 1.1arcsec,
while for the NIR images is 1.0arcsec.  
For details about the survey and data release, 
please refer to \cite{molino13}, and \cite{ch09}.
In this work, we use the public GOLD catalogs
\footnote[2]{\tt http://cosmo.iaa.es/content/ALHAMBRA-Gold-catalog},
which contains data of seven, out of the eight, ALHAMBRA fields. 
The magnitude limits of these catalogues are $ \langle m_{AB} \rangle \sim 25$ for the four blue bands and 
from $ \langle m_{AB} \rangle \sim 24.7$ mag to $23.4$ mag for the redder ones. 
The NIR limits at AB magnitudes are $J\approx$ 24 mag, $H \approx$ 23 mag, 
$K$ $\approx$ 22 mag.
In the following, the filter nomenclature presented in Table \ref{table_fname} is used
and the ALH-4 and ALH-7 fields are excluded from our analysis.
Previous authors have shown that overdensities reside in these fields 
and they might alter the redshift distribution of the selected samples 
as well as the clustering measurements (see section \S\ref{s_clustering}).
\cite{arnalte-mur14} obtained the clustering properties of ALHAMBRA galaxies 
and studied the sample variance using the seven independent ALHAMBRA fields.
They quantified the impact of individual fields on the final clustering measurements,
using the \cite{norberg11} method, they determined two ``outliers regions'', i.e. ALH-4 and ALH-7 fields.
Besides, part of the ALH-4 field spatially corresponds to the COSMOS field.
Previous works have shown that this region is dominated by
some large-scale structures (LSS), 
the most prominent are peaking at $z\sim 0.7$, and $0.9$ \citep{scoville07}. 
These structures have X-ray counterparts and a probability higher than 30$\%$ of being LSS. 
\cite{guzzo07} studied the clusters located at the center of the LSS (at $z\sim$0.7),
while \cite{finoguenov07} found diffuse X-ray emission in the most compact structures.  
There are other LSS found in the COSMOS field,
but they fall out of the redshift range of the samples studied in this paper.
%__________________________________________________________________

\section{Sample selection} \label{s_sample}
%__________________________________________________________________

To exclude the stars from the original ALHAMBRA Gold catalog we use 
the stellarity index given by SExtractor (CLASS-STAR parameter $C(K)$) and 
the statistical star/galaxy separation \cite[Section 3.6]{molino13} 
encoded in the variable Stellar-Flag ($S_{alh}$) of the catalogs.
Throughout the present paper, we define as galaxies those ALHAMBRA sources with $ C(K) \le 0.8$ and $S_{alh} \le$ 0.5.
The ALHAMBRA Gold catalog is an $F814W$ (i.e. almost an I-band)
selected catalog. 
This band was created by the ALHAMBRA team as a linear combination of
ALHAMBRA bands (see Eq. 5 in \cite{molino13}) ,
and it was used for their source detection. 
Objects with faint features in this band are not detected,
it might affect the completeness of the selected samples as will be discussed further in sections \S \ref{s_ bjgcomparison} and \S \ref{s_discussion_st}.
This catalogue is complete up to $F814W =23$ \citep{molino13},
hence we use this limit to build our complete sample.
On the other hand, the NIR completeness has been studied previously \citep{ch09}.
Using the early release of the first ALHAMBRA field,
they determined a change in the slope of the $K$-band number counts in the magnitude range $18.0<K_{Vega}<20.0$ 
and that the data is complete until $K_{VEGA}=20$ or $K_{AB}=21.8$. 
We have checked the $K$-band numbers counts of the 48 individual pointing that compose the ALHAMBRA GOLD catalog.
Every single pointing tends to falls at $K_{AB}\sim$ 22.
Hereafter, we use the $K_{AB}$=22 limit to define our complete samples. 

\subsection{Selection of star-forming galaxies} \label{ss_sfg} 

\begin{figure}[ht]
\centering
\includegraphics[scale=0.42,angle=0,trim=5.0cm -1.cm 5.cm 0cm]{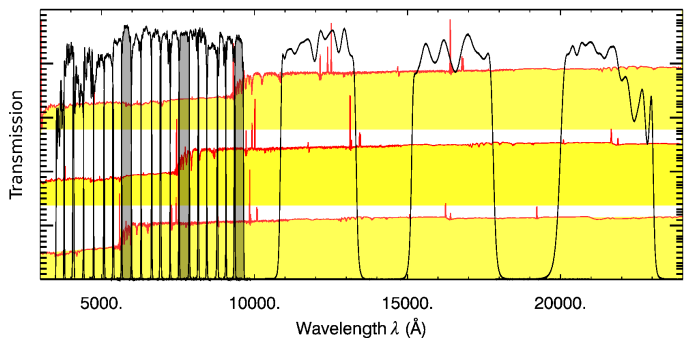}
%trim=l b r t
\caption{ALHAMBRA filter transmission curve covering from the optical to NIR.
From top to bottom, the red curves show the typical redshifted spectrum of a star-forming galaxy
at $z\sim$1.5, $z\sim$1.0 and $z\sim$0.5. 
The grey shaded areas mark the position of the ALHAMBRA filters $R_8$, $I_{14}$, and $z_{20}$, 
lightening the position of the Balmer jump.
To select the $BJG$ samples, 
we use the filter directly redwards to the gray filter to sample the ending position of the Balmer jump.
}
\label{fig1}
\end{figure}

We use the models of \cite{bc03} convolved with the ALHAMBRA filter system to define a 
two-color criteria to select star-forming galaxies analogously to the work of \cite{daddi04}.
They created the $BzK$ color-selection technique 
to cull star-forming galaxies at z>1.4.
This technique is based on the Balmer jump,
which is an indicator of recent star formation.
They used the models of \cite{bc03} to identify
the redshifted Balmer jump in the z-band,
creating the color selection criteria $BzK \equiv (z-K)_{AB}-(B-z)_{AB} = -0.2$; where
$BzK \ge -0.2$ selects star-forming galaxies at $z>1.4$.
Our approach is analogue to the $BzK$ method,
in the sense that we also use the \cite{bc03} models, % and the Balmer jump, 
whereas we record redshifted Balmer jumps in various (diverse) medium-bands 
selecting galaxies at different redshifts. 
In Fig.\ref{fig1}, this situation is illustrated, the 
typical spectrum of a star-forming galaxy redshifted to $z\sim$0.5,
$z\sim$1.0 and $z\sim$1.5 is shown. 
The ALHAMBRA bands are overplotted from the optical to NIR ranges.
We choose the $U_2$-band to sample the region before the Balmer jump instead of the $B$-band.
We select the $U_2$ band because it reaches a higher completeness level with respect to U1.
Objects that are not detected in the $U_2$ band are also considered in our selection and its magnitude limit is used.
Redder bands might sample weak features bluewards to the Balmer jump of low redshift galaxies $z < 0.4$.
The $K$ band is exactly the same used by \cite{daddi04}.
While, the z-band is replaced with the variable $X_n$-band, 
which covers the optical range from 613\AA~ to 954\AA,
i.e. $X_n$ can be any ALHAMBRA filter from $R_9$ to $z_{20}$ (see Table \ref{table_fname}).
The $X_n$-band samples the region directly red-wards to the Balmer jump,
hence the $U_2 - X_n$ color indicates the galaxy redshift range depending on the selected $n$.
On the other hand, the $X_n - K$ color registers 
the duration of the star-formation age.
The different panels in Fig.\ref{fig2} and Fig.\ref{fig3} show the theoretical evolution \citep{bc03}
of the $U_2X_nK \equiv (X_n-K)-(U_2-X_n)$ color as a function of 
redshift with $9<n<20$ ($X_n= R_9,...., z_{20})$.
The red, green and blue solid lines show the $U_2X_nK$-color evolution 
of constant star-formation rate models for ages $0.2$, $1$, 
and $2$~Gyr and reddening $E(B-V)=0.3$.
The red, green and blue dashed lines show the $U_2X_nK$-color evolution 
of passively evolving models for formation redshift of  $z_f=2,3$ and $6$. 
In each panel, the star-forming models always lie in the region $U_2X_nK>0$.
Consequently, in order to select a sample at redshift higher than $z\sim$0.5, 
a combination that involves a $X_n$ filter redder than $R_9$
must be used, as it is shown in the bottom panels of Fig. \ref{fig2} and Fig. \ref{fig3}.

\subsection{$BJG$ samples} \label{ss_bbg} 

\begin{figure*}
\centering
\includegraphics[width=.9\linewidth,angle=0,trim=2.5cm 1.cm 5.5cm 1cm, clip=true]{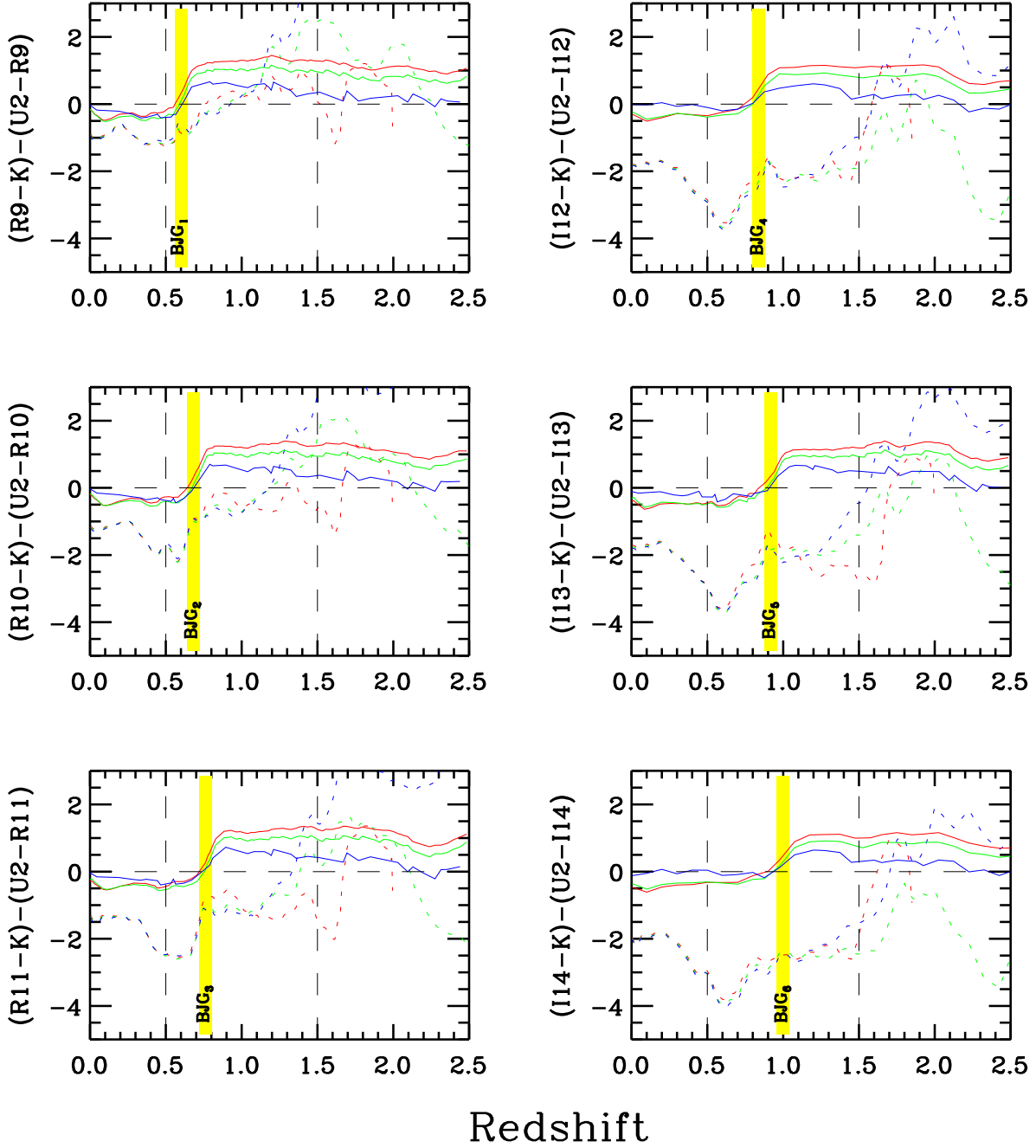}
%trim=l b r t
\caption{Criteria to select star-forming galaxies.
The solid and dashed lines show the color evolution of star-forming and passive galaxies, respectively.
The red, green, blue color of each line indicates the formation redshift of the galaxy $z_f=2,3,6$, respectively.
The horizontal dashed line indicates the color cut $U_2 X_n K \equiv (U_2-X_n)-(X_n-K) = 0$.
At each panel, the star-forming galaxies are located above this threshold.
The yellow shaded regions mark the redshift range of the selected $BJG$ samples in each color combination. Dashed vertical lines indicate the redshift 0.5 and 1.5.}
\label{fig2}
\end{figure*}

\begin{figure*}
\centering
\includegraphics[width=.9\linewidth,angle=0,trim=2.5cm 1.cm 5.5cm 1cm, clip=true]{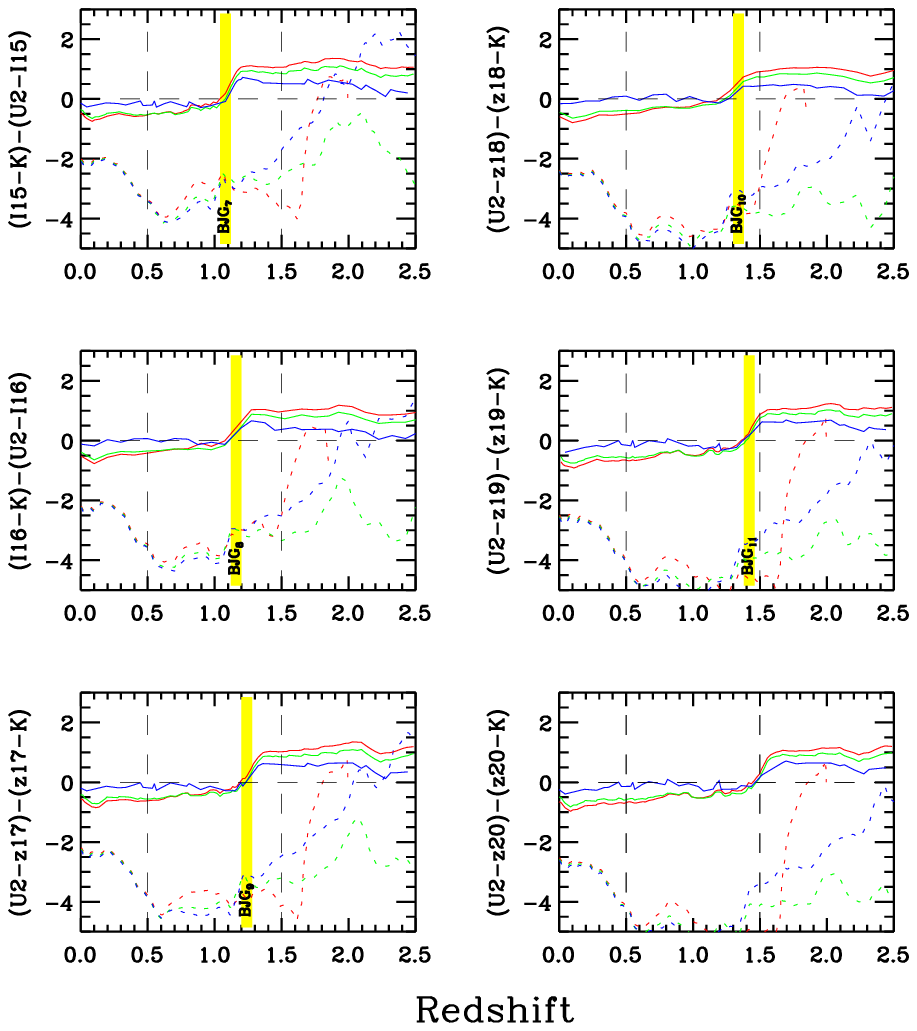}
\caption{Criteria to select star-forming galaxies.
The solid and dashed lines show the color evolution of star-forming and passive galaxies, respectively.
The red, green, blue color of each line indicates the formation redshift of the galaxy $z_f=2,3,6$, respectively.
The horizontal dashed line indicates the color cut $U_2 X_n K \equiv (U_2-X_n)-(X_n-K) = 0$.
At each panel, the star-forming galaxies are located above this threshold.
The yellow shaded regions mark the redshift range of the selected $BJG$ samples in each color combination. Dashed vertical lines indicate the redshift 0.5 and 1.5.}
\label{fig3}
\end{figure*}

We use the condition $U_2 X_n K \equiv (U_2-X_n)-(X_n-K) > 0$
and the homogeneous coverage in the optical range of the ALHAMBRA filter system,
where $n$ is ranging from $9$ to $20$ ($X_n= R_9,.. ,z_{20}$),
to select galaxies at redshift $z> 0.5, 0.6, 0.7, 0.8, 0.85, 0.95, 1.05, 1.1, 1.2, 1.25, 1.3$, respectively.
In each ALHAMBRA filter set $U_2 X_n K$, 
this color condition selects star-forming 
and passive galaxies in the \emph{wide} mentioned redshift range.
Theoretically, these selected passive galaxies lie always at higher 
redshift (i.e., $\Delta z > 0.25$) with respect to the selected star-forming galaxies.
In order to select galaxies in a \emph{narrow} redshift range,
we use more than one color condition; first the $(U_2-X_{n})-(X_{n}-K) > 0$ condition with $n=j$,
to select all galaxies above certain redshift $z_j$
and secondly we subtract the higher-redshift 
samples, selected with $n \ge j+1$ (passive and star-forming), from the first galaxy selection. 
For example, our lowest redshift sample was selected by imposing the condition
$U_2R_9K = 0$ (i.e. select galaxies at $z > 0.5$)
and subtracting the galaxy samples selected by $U_2X_{n}K > 0$ with $n \ge 10$,
which select all galaxies at $z > 0.6$.
In this way, only the star-forming galaxies in the 
redshift range $0.5<z<0.6$ were selected (see yellow shaded region in Fig.\ref{fig2} and Fig. \ref{fig3}).
Following this method and using the homogeneous separation between each ALHAMBRA band, 
we select eleven star-forming galaxy samples peaking at 
$z\sim 0.55, 0.65, 0.75, 0.8, 0.9,  1.0, 1.05, 1.15, 1,2, 1.25$, and $z\sim1.4$.
It by culling the galaxies that satisfy the star-forming criteria 
$U_2 X_n K \equiv (U_2-X_n)-(X_n-K) > 0$, 
where $X_n= R_9,..., z_{19}$
and ``cleaned'' of higher redshift galaxies by subtracting the samples that satisfy
$U_2X_{n}K> 0$, with $n \ge 10, ..., 20$, respectively.
Table \ref{table_sample} summarizes the properties 
of the selected samples using the method described above.
We ensure that all samples are roughly independent of each other
by removing the high redshift samples, whose are selected using the $X_n$ filters
with $n \geq j+1$ (see column 3 of Table \ref{table_sample}), 
from the sample selected with the filter $X_{n=j}$.

\begin{table}
\caption{Properties of the $BJG$ selected samples.}
\begin{tabular}{lcccc}
\hline\hline                
Sample &Initial set &Clean sets &$N$   & <z>\\ 
Name   &$U_2X_nK$ &$U_2X_nK$  	 & 	 &Two-color\\      
\hline \hline
\noalign{\smallskip}
$BJG_1$	&$R_9$		&$n\ge 10$	&5489	& 0.5-0.6\\
\noalign{\smallskip}	 
$BJG_2$	&$R_{10}$	&$n\ge 11$	&5174	& 0.6-0.7\\
\noalign{\smallskip} 
$BJG_3$	&$R_{11}$	&$n\ge 12$	&4497   & 0.7-0.8\\
\noalign{\smallskip}
$BJG_4$	&$I_{12}$	&$n\ge 13$	&4012	& 0.8-0.85\\
\noalign{\smallskip}
$BJG_5$	&$I_{13}$	&$n\ge 14$	&3550	& 0.85-0.95\\
\noalign{\smallskip}
$BJG_6$	&$I_{14}$	&$n\ge 15$	&2878	& 0.95-1.05\\
\noalign{\smallskip}
$BJG_7$	&$I_{15}$	&$n\ge 16$	&2231	& 1.05-1.1\\
\noalign{\smallskip}
$BJG_8$	&$I_{16}$	&$n\ge 17$	&2325	& 1.1-1.2\\
\noalign{\smallskip}
$BJG_9$	&$z_{17}$	&$n\ge 18$	&2058	& 1.2-1.25\\
\noalign{\smallskip}
$BJG_{10}$ &$z_{18}$	&$n\ge 19$	&2140   & 1.25-1.3\\
\noalign{\smallskip}
$BJG_{11}$ &$z_{19}$	&$n=20$		&3391	& 1.3-1.5\\
\hline\hline                 
\noalign{\smallskip}	 

\end{tabular}
\\
{\bf Notes.}
Col. 1, Name of the selected sample;
Col. 2, Initial filter set used for selection;
Col. 3, Second filter sets used to subtract higher redshift galaxies from the initial sample;
Col. 4, Number of galaxies selected.
Col. 5, Expected redshift range of the sample.
 \\
\label{table_sample}
 \end{table}
  
%__________________________________________________________________

\section{Physical properties of the $BJG$ samples}\label{s_pp} 

%__________________________________________________________________

In this section, we aim to characterize each $BJG$ sample, 
providing a mean characteristic value of its redshift, 
absolute magnitude, stellar mass, age, star formation rate, etc.

\begin{figure}[h]
\centering
\includegraphics[scale=1.0,angle=0,trim=20.0cm 0.cm 20.cm 0cm]{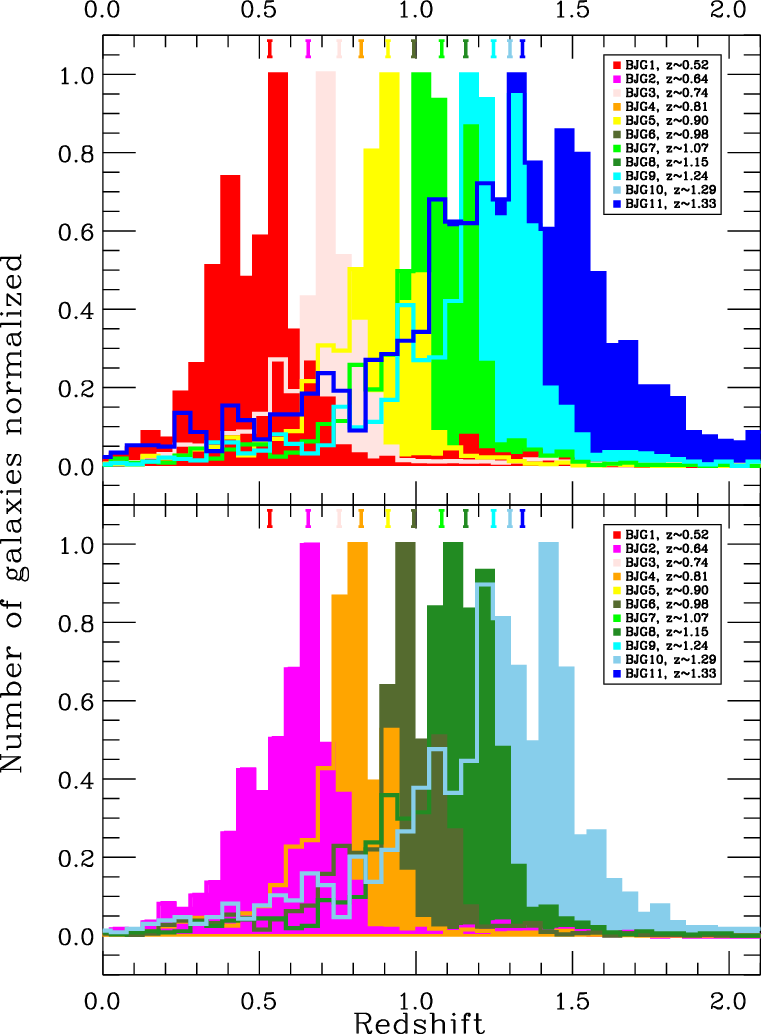}
%trim=l b r t
\caption{Photometric redshift distribution of the eleven selected samples.
For a better visualization of the distribution tails,
the upper panel shows the BPZ distribution of the 
$BJG_1,BJG_3,BJG_5,BJG_7,BJG_9$, and $BJG_{11}$ samples,
while the lower panel presents the 
$BJG_2,BJG_4,BJG_6,BJG_8$, and $BJG_{10}$ samples.
The vertical colored lines show the median of the zBPZ distribution for the selected samples. 
}
\label{fig_bpz}
\end{figure}

\subsection{Photometric redshifts}\label{s_bpz}

We use the Bayesian photometric redshift (zBPZ) published in the 
Gold ALHAMBRA catalogs \citep{molino13} to verify our selection method.
The  BPZ code was optimized to determine photometric redshifts,
for details on the zBPZ calculations see \cite{molino13} and \cite{benitez09}.
Fig.\ref{fig_bpz} shows the zBPZ distribution of the eleven $BJG$ samples.
For a better visualization of the distribution tails,
the upper panel shows the BPZ distribution of the 
$BJG_1,BJG_3,BJG_5,BJG_7,BJG_9$, and $BJG_{11}$ samples,
while the lower panel presents the 
$BJG_2,BJG_4,BJG_6,BJG_8$, and $BJG_{10}$ samples.
The vertical colored lines show the median of the zBPZ distribution for the selected samples. 
The median of zBPZ distribution, reported on Table \ref{table_pp}, clearly agree 
with the median of the expected redshift range estimated
via the two-color selection criteria that are reported on Table \ref{table_sample}.

\subsection{SED fitting}\label{s_isedfit} 

For each galaxy of the $BJG$ samples, we fit the 20 optical bands plus the three NIR bands,
using the Bayesian SED modeling code {\tt iSEDfit} \citep{moustakas13}.
Once we fix the redshift to the best-fit value given by BPZ in the ALHAMBRA catalog \citep{molino13},
{\tt iSEDfit} calculates the marginalized posterior probability distributions for the physical parameters,
in certain model space that is previously defined by the user.
Using a Monte Carlo technique, we generate 20,000 model SEDs with \emph{delayed} star-formation histories
SFH$\sim te^{-t/\tau}$, where $\tau$ is the star formation timescale.
The SEDs were computed employing the Flexible Stellar Population Synthesis models \citep[FSPS, v~2.4;][]{conroy09, conroy10}
based on the {\sc miles} \citep{sanchez-blazquez06} and Basel \citep{basel1,basel2,basel3} stellar libraries.
We assume a \cite{chabrier03} initial mass function from $0.1-100~M_{\odot}$,
and a time-dependent attenuation curve of \cite{charlot00}.
We adopt uniform priors on stellar metallicity $Z/Z_{\odot}\in[0.04,1.0]$,
galaxy age $t\in[0.01,age(z_{BPZ})]$~Gyr,
rest-frame $V$-band attenuation $A_{V}\in[0-3]$~mag,
and star formation timescale $\tau\in[0.01,age(z_{BPZ})]$~Gyr, 
where $age(z_{BPZ})$ is the age of the Universe at each galaxy's photometric redshift.
Figures~\ref{fig_isedfit} and \ref{fig_isedfit2} 
show the SEDs of a galaxy randomly picked out of each $BJG$ sample. 
The filled green dots show the photometric data,
the red line shows the model that minimizes the $\chi^2$ using {\tt iSEDfit},
the black squares mark the convolution between this model and the ALHAMBRA filters.
The blue shading shows the universe of models, generated by {\tt iSEDfit} using the previously described priors,
scaled by their reduced $\chi^2$. The color bar indicates the reduced $\chi^2$ scale.

In Table \ref{table_pp}, for each $BJG$ sample,
the median value of the sum of the posterior probability distributions for some physical properties are reported.
The uncertainties indicate the 1-$\sigma$ confidence level, to account for asymmetric distributions we determine the percentiles 16 and 84.

\begin{table}
\caption{Median physical properties of the $BJG$ samples.}
\begin{tabular}{cccccc}
\hline\hline
Name & N &zBPZ  &<$M_*$>    &SFR 		&Age  \\
     &   &       &M$_\odot$  &M$_\odot yr^{-1}$  &[Gyr]  \\      
\hline\hline 
\noalign{\smallskip}	 
      $BJG_{ 1}$  &5489 &$0.52^{+  0.16}_{-0.14 }$ &$10.15^{+  0.11}_{-0.13 }$ &$0.08^{+  0.22}_{-0.23 }$ &$5.52^{+  1.62}_{-1.87}$  \\
\noalign{\smallskip}
      $BJG_{ 2}$  &5174 &$0.64^{+  0.11}_{-0.18 }$ &$10.25^{+  0.12}_{-0.13 }$ &$0.27^{+  0.23}_{-0.22 }$ &$5.13^{+  1.50}_{-1.76}$  \\
\noalign{\smallskip}
      $BJG_{ 3}$  &4497 &$0.74^{+  0.10}_{-0.17 }$ &$10.30^{+  0.13}_{-0.14 }$ &$0.38^{+  0.23}_{-0.24 }$ &$4.83^{+  1.40}_{-1.68}$  \\
\noalign{\smallskip}
      $BJG_{ 4}$  &4012 &$0.81^{+  0.12}_{-0.14 }$ &$10.37^{+  0.13}_{-0.15 }$ &$0.42^{+  0.25}_{-0.26 }$ &$4.67^{+  1.29}_{-1.62}$  \\
\noalign{\smallskip}
      $BJG_{ 5}$  &3550 &$0.90^{+  0.12}_{-0.18 }$ &$10.36^{+  0.14}_{-0.16 }$ &$0.49^{+  0.24}_{-0.25 }$ &$4.37^{+  1.24}_{-1.54}$  \\
\noalign{\smallskip}
      $BJG_{ 6}$  &2878 &$0.98^{+  0.12}_{-0.19 }$ &$10.41^{+  0.15}_{-0.17 }$ &$0.52^{+  0.25}_{-0.27 }$ &$4.12^{+  1.15}_{-1.46}$  \\
\noalign{\smallskip}
      $BJG_{ 7}$  &2231 &$1.07^{+  0.12}_{-0.20 }$ &$10.46^{+  0.15}_{-0.17 }$ &$0.63^{+  0.25}_{-0.26 }$ &$3.83^{+  1.09}_{-1.40}$  \\
\noalign{\smallskip}
      $BJG_{ 8}$  &2325 &$1.15^{+  0.12}_{-0.21 }$ &$10.45^{+  0.16}_{-0.19 }$ &$0.69^{+  0.26}_{-0.27 }$ &$3.63^{+  1.05}_{-1.36}$  \\
\noalign{\smallskip}
      $BJG_{ 9}$  &2058 &$1.24^{+  0.14}_{-0.25 }$ &$10.45^{+  0.18}_{-0.20 }$ &$0.72^{+  0.25}_{-0.25 }$ &$3.41^{+  1.01}_{-1.32}$  \\
\noalign{\smallskip}
      $BJG_{10}$  &2140 &$1.29^{+  0.21}_{-0.33 }$ &$10.38^{+  0.19}_{-0.23 }$ &$0.72^{+  0.24}_{-0.22 }$ &$3.16^{+  1.05}_{-1.29}$  \\
\noalign{\smallskip}
      $BJG_{11}$  &3391 &$1.33^{+  0.28}_{-0.42 }$ &$10.34^{+  0.20}_{-0.24 }$ &$0.72^{+  0.23}_{-0.21 }$ &$2.99^{+  1.05}_{-1.27}$  \\
\hline\hline
\noalign{\smallskip}	 
\end{tabular}
\\
{\bf Notes.} 
Col. 1, Name of the selected sample;
Col. 2, number of galaxies selected;
Col. 3, BPZ photometric redshift;
Col. 4, logarithm of the stellar mass (Chabrier IMF);
Col. 5, logarithm of the star formation rate;
Col. 6, galaxy age.
\\
\label{table_pp}
\end{table}

\subsection{Comparison between $BJG$ samples}
\label{s_ bjgcomparison}
\begin{figure}
\centering
\includegraphics[width=1.\linewidth,angle=0,trim=0.5cm .5cm .5cm .5cm, clip=true]{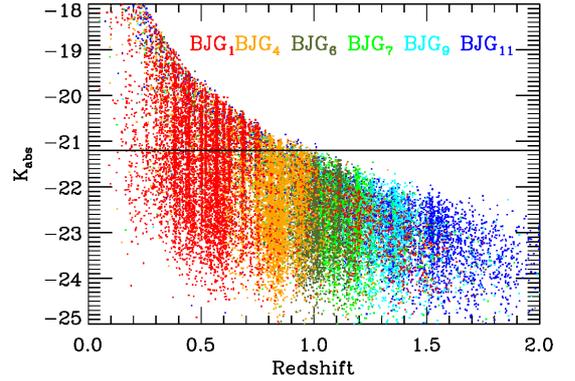}
%trim=l b r t
\caption{Absolute magnitude as a function of redshift.
The red, orange, green, cyan, light blue, and blue dots indicates the galaxies in
the $BJG_1$, $BJG_4$, $BJG_6$, $BJG_7$, $BJG_9$, and $BJG_{11}$ 
samples with apparent magnitude brighter than $K=22$, respectively.
The black solid line shows the absolute magnitude cut $K_{abs}=-21.2$.
}
\label{fig_completitud}
\end{figure}

\begin{figure*}
\centering
\includegraphics[width=.8\linewidth,angle=0,trim=1.cm 1.cm .5cm 1cm, clip=true]{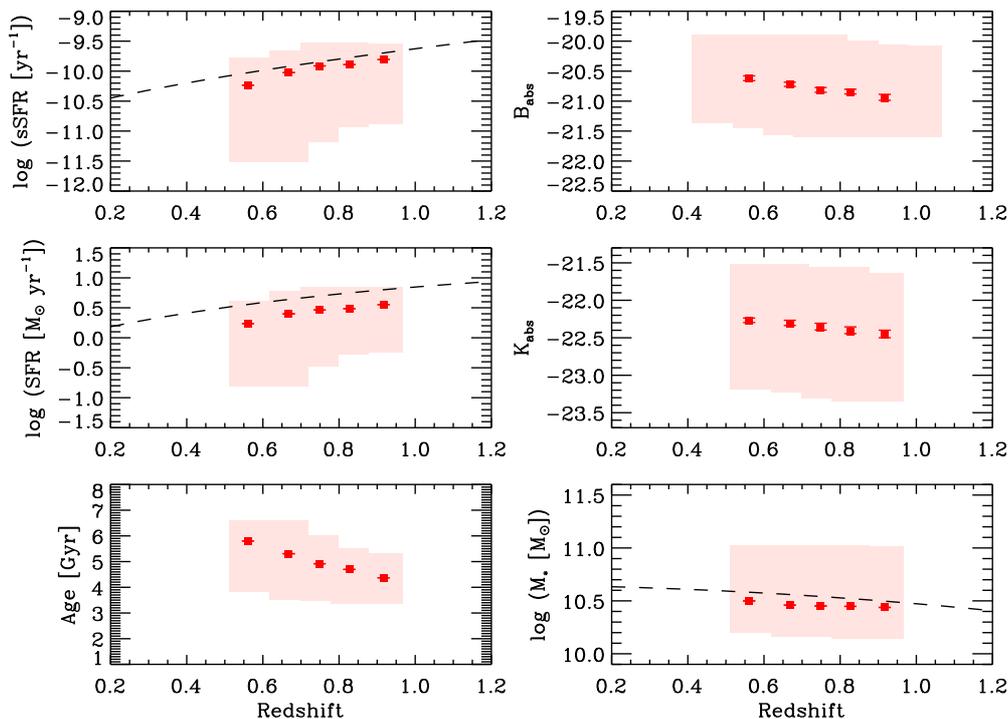}
\caption{Evolution of the physical properties of the $BJGK_{n<6}$ samples. 
The panels show the median of the probability distribution of, from up to down and left to right, 
the specific star formation rate, star formation rate, age, $B-$band luminosity, 
$K-$band luminosity, and stellar mass, respectively.
Red squares show the evolution of $BJGK_{n<6}$ galaxies brighter than $K=-21.2$.
The colored areas indicates the dispersion of each sample,
while the error bars show the median error of the physical properties determined by {\tt iSEDfit}. 
Dashed lines show the fit to the CANDELS analysis taken from \cite{vandokkum13} 
for galaxies of stellar masses similar to the Milky Way.}
\label{fig_ppevol}
\end{figure*}

In this subsection, we aim to compare the properties, derived from SED fitting, 
of the $BJG$ samples for galaxies of similar $K$-band absolute magnitude, i.e. similar stellar mass.
Therefore, in addition to restricting the $BJG$ samples within the magnitude survey limit
(see section \S\ref{s_sample}) we also apply an absolute magnitude limit for all samples.
In the following, we search for the most appropriate absolute magnitude limit 
that allows us to build these samples.
Fig.\ref{fig_completitud} shows the $K$-band absolute magnitude,
obtained via {\tt iSEDfit}, as a function of BPZ redshift.
The red, orange, green, cyan, light blue and blue dots indicate the galaxies
in the $BJG_1$, $BJG_4$, $BJG_6$, $BJG_7$, $BJG_9$, and $BJG_{11}$ 
samples with apparent magnitude brighter than the magnitude survey limit $K=22$, respectively.
We can note that {\it partly} due to Malquist bias,
the farthest sample $BJG_{11}$ is roughly complete only until $K_{abs}=-22.5$ (blue dots).
However, by choosing this bright absolute limit to compare all $BJG$ samples 
we restrict our study only to the most massive and bright objects
as well as enormously reduce the statistics for each sample, specially at low redshift.
Hence, we decide to study all objects brighter than $K_{abs} = -21.2$,
which correspond to the absolute limit where the $BJG_5$ sample at $z\sim$1
is complete and comparable, in terms of $K_{abs}$ luminosity, 
to the other lower redshift $BJG_{n<6}$ samples.
In Fig.\ref{fig_completitud} the black solid line shows the absolute magnitude limit $K_{abs}=-21.2$.
We have chosen the $z\sim1$ limit because the ALHAMBRA Gold data is an $F814W$ (i.e. almost an I-band) selected catalog and
thus less sensitive to galaxies at $z\geq$1 with a pronounced Balmer jump.
For galaxies at $z\geq$ 1, the spectral region directly bluewards of the Balmer jump is barely or not detected in the $F814W$
because the Balmer jump falls in the $I_{13}$ and the region directly bluewards to it falls in the $R_{12}$ band.
Since the $F814W$ image detection is a linear combination involving the $R_{12}$ band, this kind of galaxies are barely or not detected as sources in the final catalog.
We expect a deeper selection of galaxies with a pronounced Balmer jump at $z<$1.
Said that, the $BJG$ samples at $z\leqslant1$ selected from $X_{n \leq 13}$ bands 
are optimized to be complete according to the $F814W$ selection, whereas the samples at $z\geq$1 are incomplete.
The level of incompleteness of the $BJG_{n\geq6}$ samples is difficult to quantify because there may be many effects 
working together (e.g. the $F814W$ is a linear combination of ALHAMBRA bands, 
undetectable bluewards regions of the Balmer jump due to intrinsic galaxy properties such as redshift, dust excess, etc.).
To minimize the incompleteness of our samples
in the comparison of the physical properties at different redshifts, 
in the following we re-define the first five $BJG$ samples selecting 
all galaxies brighter than $K_{abs}=-21.2$. 
Furthermore, the $BJG_{n}$ samples, with $n\ge6$, are not considered in the following analysis.  
Table~\ref{table_ppK} presents the properties of the final sample definition ($BJGK_n$, with $n<6$), 
which takes into account the absolute luminosity cut ($K_{abs}=-21.2$).
In Fig.\ref{fig_ppevol}, the comparison of the physical properties 
of galaxies with similar $K-$band luminosity is shown.
Red squares show the properties of the $BJGK_n$ (with $n<6$),
the dashed lines show the results of \cite{vandokkum13} 
for galaxies of similar stellar mass.
The colored areas indicate the dispersion of each sample,
while the error bars show the median error of the physical properties derived  via {\tt iSEDfit}. 
In section \S \ref{s_discussion_pp} these results are discussed.

\begin{table}
\caption{Median physical properties of the $BJGK$ samples considering the absolute magnitude limit $K=-21.2$.}
\begin{tabular}{cccccc}
\hline
\hline
Name &N  &zBPZ    &<$M_*$>    &SFR 		&Age \\
     & &	  &M$_\odot$   &M$_\odot yr^{-1}$ &[Gyr]\\      
\hline\hline 
\noalign{\smallskip}	 
$BJGK_1$	&3322 &$0.56^{+0.18}_{-0.14}$ &$10.50^{+0.10}_{-0.11}$ &$0.24^{+0.29}_{-0.30}$ &$5.80^{+1.43}_{-1.87 }$  \\
\noalign{\smallskip} 
$BJGK_2$	&3778 &$0.67^{+0.11}_{-0.16}$ &$10.46^{+0.11}_{-0.12}$ &$0.40^{+0.27}_{-0.26}$ &$5.31^{+1.37}_{-1.74 }$  \\
\noalign{\smallskip} 
$BJGK_3$	&3576 &$0.75^{+0.10}_{-0.12}$ &$10.45^{+0.12}_{-0.13}$ &$0.46^{+0.26}_{-0.26}$ &$4.92^{+1.31}_{-1.67 }$  \\
\noalign{\smallskip} 
$BJGK_4$	&3480 &$0.83^{+0.11}_{-0.09}$ &$10.45^{+0.13}_{-0.14}$ &$0.49^{+0.27}_{-0.28}$ &$4.69^{+1.24}_{-1.60 }$  \\
\noalign{\smallskip} 
$BJGK_5$	&3150 &$0.92^{+0.10}_{-0.12}$ &$10.44^{+0.14}_{-0.15}$ &$0.55^{+0.26}_{-0.26}$ &$4.37^{+1.19}_{-1.52 }$  \\
\noalign{\smallskip} 
\hline\hline
\noalign{\smallskip}	 

\end{tabular}
\\
{\bf Notes.} 
Col. 1, Name of the selected sample;
Col. 2, number of galaxies selected from the $BJG$ samples after the $K$-band absolute cut;
Col. 3, BPZ photometric redshift;
Col. 4, stellar mass (Chabrier IMF);
Col. 5, star formation rate;
Col. 6, galaxy age.
\\
\label{table_ppK}
\end{table}

%__________________________________________________________________

\section{Halo masses through clustering analysis}\label{s_clustering}

%__________________________________________________________________

In hierarchical clustering, structures build up in time from small 
density fluctuations. Small structures agglomerate to build large 
structures. 
Dark matter haloes are biased tracers of the underlying matter density field.
Massive haloes lie in higher and rarer density peaks and are more clustered than lower mass haloes.
If a galaxy population is hosted by haloes of a given mass, 
then the clustering amplitude of the galaxy population, 
compared to the expected dark matter clustering at the same redshift, 
can be used to derive the typical halo mass corresponding to that galaxy population. 
This is encapsulated in the bias parameter, $b$, defined as $b^2=\xi_{gal}(r)/\xi_{DM}(r)$,
where $\xi_{gal}$ and $\xi_{DM}$ are the 2-point spatial correlation function for 
galaxies and dark matter, respectively.
The correlation function at a certain redshift, $\xi(r,z)$,
can be characterized with a power law and a correlation length, $r_0$, 
through $\xi(r,z)=(r/r_0(z))^\gamma$, where $\gamma$ is the power index.
For sparse samples and at small scales this representation is typically used.
In larger surveys and simulations the correlation function is usually
modelled combining terms for galaxies in the same halo 
and in different ones \citep{zehavi11,contreras13}.

In the following, we measure the bias and 
masses of the halos that host the $BJGK_{n<6}$ samples.
Specifically, for each $BJGK_{n<6}$ sample,
we calculate the 2-point angular correlation 
function, and then, using Limbers’ de-projection,
given a redshift distribution and a cosmological model,
we calculate the correlation length.
The bias parameter follows from $r_0$ through $b^2\simeq \sigma_{8,gal}^2(r_0,z)/\sigma_{8,DM}^2(z)$,
where $\sigma_{8,gal}$ ($\sigma_{8,DM}$) is the root mean square fluctuation amplitude 
in 8$h^{-1}$~Mpc spheres of galaxies (dark matter).
The halo mass is obtained from the bias by using the models of \cite{sheth_mo_tormen01}.
We have chosen this procedure and the 8~$h^{-1}$Mpc scale, 
in order to directly compare our results with previous works
that connected progenitors and descendants of samples of star-forming galaxies
\citep[see Fig. \ref{fig_bias},][]{adelberger05,ouchi05,kashikawa06,gawiser07,hildebrandt07,hayashi07,blanc08,hartley08,yoshida08,guaita10,mccracken10,lin12}.
More detailed treatments, such as considering both halo terms and different scales,
are beyond the scope of this paper.

\begin{table*} \label{table_clustering}
\centering
\caption{Clustering properties of the $BJGK$ samples.}
\begin{tabular}{lccccccc}
\hline
\hline
Name   &$N_{gal}$  &<z> &<$K_{abs}$> &$A_w$ &$r_0$ &Bias 	&log $M_h$\\
       & 	  	   & BPZ     & mag	 & $10^{-3}$ &[$h^{-1}$ Mpc] &(r=8~$h^{-1}$~Mpc)	&[$h^{-1} M_{\sun}$]  \\      
\hline
\hline 
\noalign{\smallskip}	 
$BJGK_1$	   &3332	&$0.56^{+0.2}_{-0.1}$	&-22.3	&4.2$\pm$0.2	& 4.67$\pm$0.33	& 1.36$\pm$0.09 &$ 12.71^{+0.13}_{-0.15}$\\
\noalign{\smallskip}	 
$BJGK_2$	   &3778	&$0.67^{+0.1}_{-0.2}$	&-22.3	&4.3$\pm$0.5	& 4.58$\pm$0.39	& 1.41$\pm$0.11 &$ 12.66^{+0.15}_{-0.17}$\\
\noalign{\smallskip}
$BJGK_3$	   &3576	&$0.75^{+0.1}_{-0.1}$	&-22.4	&4.0$\pm$0.2	& 3.79$\pm$0.25	& 1.24$\pm$0.07 &$ 12.27^{+0.14}_{-0.16}$\\
\noalign{\smallskip}
$BJGK_4$	   &3480	&$0.83^{+0.1}_{-0.1}$	&-22.4	&4.2$\pm$0.1	& 4.15$\pm$0.23	& 1.40$\pm$0.07 &$ 12.45^{+0.10}_{-0.12}$\\
\noalign{\smallskip}
$BJGK_5$	   &3150	&$0.92^{+0.1}_{-0.1}$	&-22.5	&3.7$\pm$0.2	& 4.01$\pm$0.23	& 1.40$\pm$0.07 &$ 12.37^{+0.11}_{-0.12}$\\
\noalign{\smallskip}
\hline\hline
\noalign{\smallskip}	 

\end{tabular}
\\
\justify
{\bf Notes.} 
Col. 1, Name of the selected sample;
Col. 2, number of galaxies selected from the $BJG$ samples after the $K$-band absolute cut and image mask;
Col. 3, BPZ photometric redshift;
Col. 4, medium $K$-band absolute magnitude;
Col. 5, amplitude of correlation;
Col. 6, correlation length;
Col. 7, bias factor calculated with the variance at 8~$h^{-1}$~Mpc.
Col. 8, logarithm of the halo mass in units of [$h^{-1} M_{\sun}$].
\\
\end{table*}

\subsection{Angular correlation and correlation length}
To determine the angular correlation function, 
we use the Landy and Szalay (1993) prescription 
\begin{equation}
 w(\theta)=(N_{gg} - 2N_{gr} + N_{rr})/N_{rr} ,
 \end{equation}
where $N_{gg}$ and $N_{rr}$ are the number of pairs at a separation $\theta$ of 
galaxies in the catalog and points in a random catalog with the same 
layout as the galaxy sample, respectively.
$N_{gr}$ is the cross number 
of points between the galaxy and random distributions.
For our calculations we consider five ALHAMBRA fields (see section \S \ref{s_alh}) that encompass 36 pointing catalogs.
Errors were estimated by a jackknife method,
which has been shown as a robust error estimator \citep{zehavi02,cabre07},
although it can overestimate the variance on small scales \citep{norberg09}.
We have calculated the angular correlation function 36 times,
each time eliminating one catalog out of the 36 available.
The uncertainty is estimated as the variance of $w(\theta)$.
We  follow  the  method  of \cite{infante94,quadri07} to  correct
for the integral constraint.
As mentioned above, to calculate $r_0$ 
we use the power law approximation 
$\xi(r,z)=(r(z)/r_0(z))^{-\gamma}$ 
that depends on the spatial separation and redshift.
Likewise, the 2-point angular function turns out to be 
$w(\theta,z)=A_w(z)\theta^{(1-\gamma)}$,
where $A_w(z)$ is the angular amplitude.
By fitting the correlation function to this power law,
between 0.005 and 0.2 degree, the $A_w(z)$ is inferred.
The angular amplitudes are reported in Table \ref{table_clustering}.
To calculate the spatial from the angular function, 
we use the Limber’s (1953) inversion, 
which requires a redshift distribution $N(z)$, 
and assume a cosmological model. 
Limber’s inversion involves solving the following integral \citep{kovac07}, 
\begin{equation}
r_0^\gamma=\frac{ A_w (c/H_0) {[\int {N(z) dz}]}^2}{C_\gamma \int{F(z) D_\theta^{1-\gamma}(z) N^2(z) g_d(z) dz} } ,
\end{equation}
where the cosmology plays the role in the Hubble parameter $H_0$,
$g_d(z)= (1+z)^2\sqrt{1+\Omega_m z+\Omega_\Lambda[(1+z)^{-2}-1]}$,
and in the angular diameter distance $D_\theta$.
The parameter $C_\gamma$ depends on the power index such that 
$C_\gamma=\Gamma(0.5)\frac{\Gamma[0.5(\gamma-1)]}{\Gamma[0.5\gamma]}$.
In order to be consistent internally and compare with other works \citep{adelberger05,ouchi05,kashikawa06,hayashi07,hildebrandt07,gawiser07,blanc08,hartley08,yoshida08,guaita10,mccracken10,lin12},
the value of $\gamma$ was fixed to the canonical $\gamma=1.8$.
This value is fully justified by experiments 
where $\gamma$ was left free and is consistent with the clustering of
luminosity-selected ALHAMBRA samples \citep{arnalte-mur14}.
$F(z)$ accounts for the redshift evolution of
the correlation function, where $F(z)=(1+z)^{-(3+\epsilon)}$,
and we use $\epsilon=-1.2$ that corresponds to the value adopted for
a constant clustering in comoving coordinates \citep{quadri07}.
For the calculation of the correlation length error,
the error contribution from the amplitude of the angular correlation function,
the Poissonian errors ($\sim \sqrt{N(z)}$),
and the effects of the photometric redshift error in shifting and broadening N(z) are taken into account.

\subsection{Bias and mass measurements} \label{ss_biashalomass}

In turn we estimate the bias parameter, $b$, from the correlation length.
As pointed out above, $b$ is related to $r_0$ through the spatial correlation function by 
$b^2 \approx \xi_{gal}(r,z)/\xi_{DM}(r,z)$ or $b^2 \approx \sigma^2_{gal}(r,z)/\sigma^2_{DM}(r,z)$. 
The numerator $\xi_{gal}(r,z) \approx J_2 \sigma^2_{gal}(r,z)$
is taken from \cite{peebles80}{(Eq. 59.3)}, where $J_2$ is a 
parameter defined in terms of $\gamma$ as  $J_2=72/[(3-\gamma)(4-\gamma)(6-\gamma)2^\gamma]$.
We fix the spatial scale at 8 $h^{-1}$ Mpc such 
that $\xi_{gal}(8,z)=(r_0(z)/8 h^{-1}Mpc)^\gamma$.
The evolution of the dark matter density variance 
in a comoving sphere of radius $8h^{-1}$~Mpc is $\sigma_{DM}^2(8,z)=\sigma_8D(z)$,
where $D(z)$ is the linear growth factor at redshift z.
The bias measured for each $BJGK$ sample, at the scale of $8h^{-1}$~Mpc, 
is reported in Table~\ref{table_clustering}.

\begin{figure*}
\centering
\includegraphics[width=.8\linewidth,angle=0,trim=2.cm 0.5cm .5cm 1cm, clip=true]{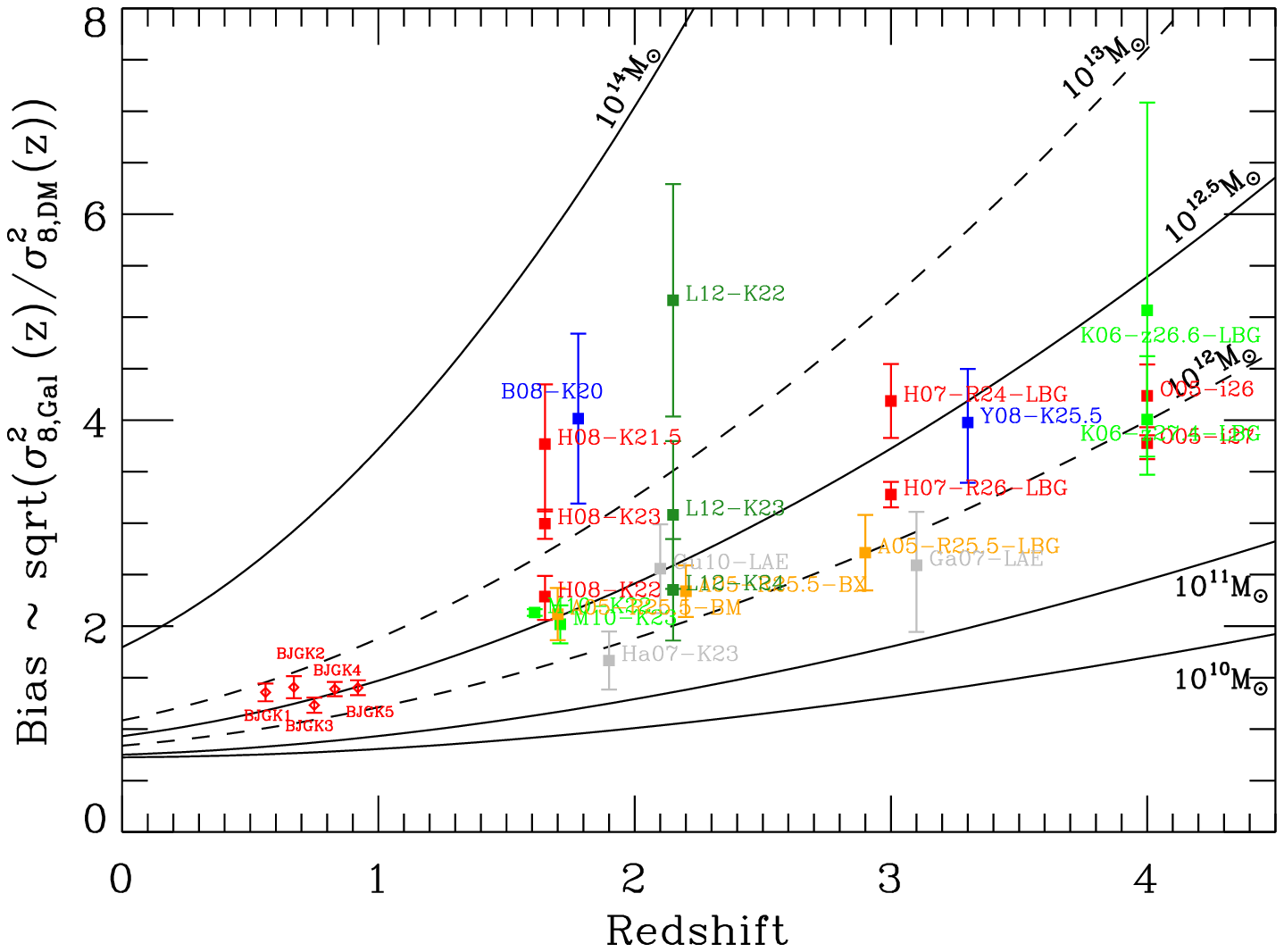}
\caption{Evolution of the bias factor, calculated with the variance at 8~$h^{-1}$Mpc,
as function of redshift for samples of star-forming galaxies.
Red diamonds indicates the $BJGK_{n<6}$ samples. 
At $z\sim$2, the grey, blue, red, green, and light green squares 
show the $sBzK$ samples of \cite{hayashi07,blanc08,hartley08,lin12}, and \cite{mccracken10}, respectively.
Yellow squares show the $BM$, $BX$ selected galaxies of \cite{adelberger05}.
At $z\sim$3, the blue, green and red squares indicates the $LBG$s selected by 
\cite{yoshida08,kashikawa06,hildebrandt07,ouchi05}, respectively.
Gray squares show the Ly-$\alpha$ emitters selected by \cite{guaita10,gawiser07}.
The lines show the bias evolution for different halo masses \citep{sheth_mo_tormen01}, 
whose are indicated below of each line in the right side of the panel.
}
\label{fig_bias}
\end{figure*}

\begin{figure}
\centering
\includegraphics[scale=0.7,angle=0,trim=0.0cm .0cm 0.cm 0.cm]{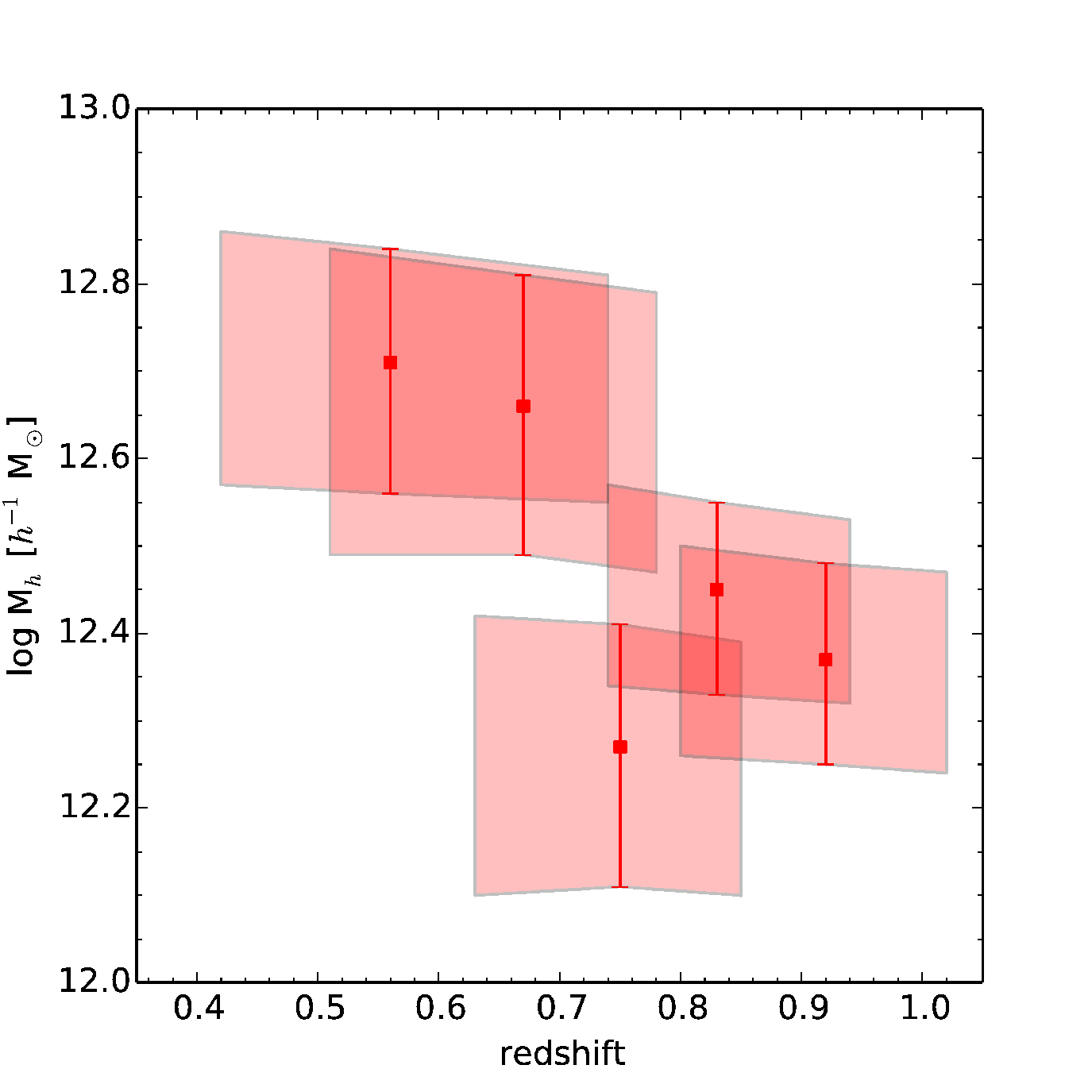}
%trim=l b r t
\caption{Halo mass, calculated with the variance at 8~$h^{-1}$Mpc, 
as function of redshift for the $BJGK_{n<6}$ samples.
Squares show the median values of the halo mass, polygons show a conservative limit,
drawing the mass limits taking into account the $1\sigma$ deviation of the redshift and bias.
}
\label{fig_ivan}
\end{figure}

\begin{figure*}
\centering
%trim=l b r t
\includegraphics[scale=1.,angle=0,trim=5cm 2cm 0cm 0cm]{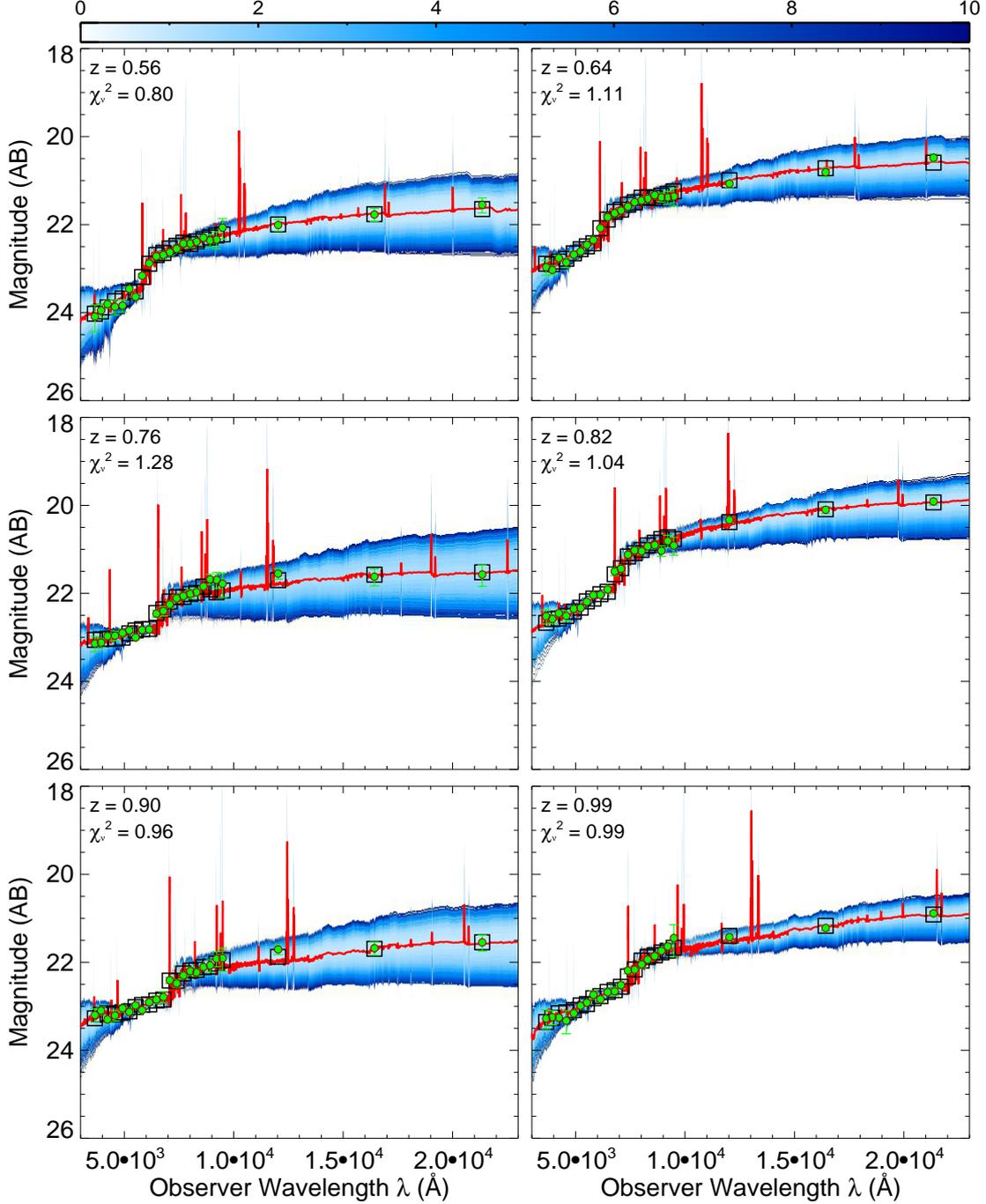}
\caption{Spectral Energy Distribution of a star-forming galaxy in the ALHAMBRA survey.
From left to right and up to bottom, the panel shows the SED of a galaxy randomly picked out 
from the $BJG_1$, $BJG_2$, $BJG_3$, $BJG_4$, $BJG_5$, and $BJG_6$ samples, respectively.
The filled green dots show the ALHAMBRA photometric data.
The red line shows the model that minimizes the $\chi^2$ found by {\tt iSEDfit}, 
the minimum reduced $\chi^2$ is indicated in the upper left corner,
beside the BPZ photometric redshift.
The black squares mark the ALHAMBRA photometry of the best model.
The blue shading shows the Universe of models, generated by {\tt iSEDfit} (see Section \S\ref{s_isedfit}),
scaled by their reduced $\chi^2$. The color bar indicates the reduced  $\chi^2$ scale.
}
\label{fig_isedfit}
\end{figure*}

\begin{figure*}
\centering
%trim=l b r t
\includegraphics[scale=1.,angle=0,trim=5cm 2cm 0cm 0cm]{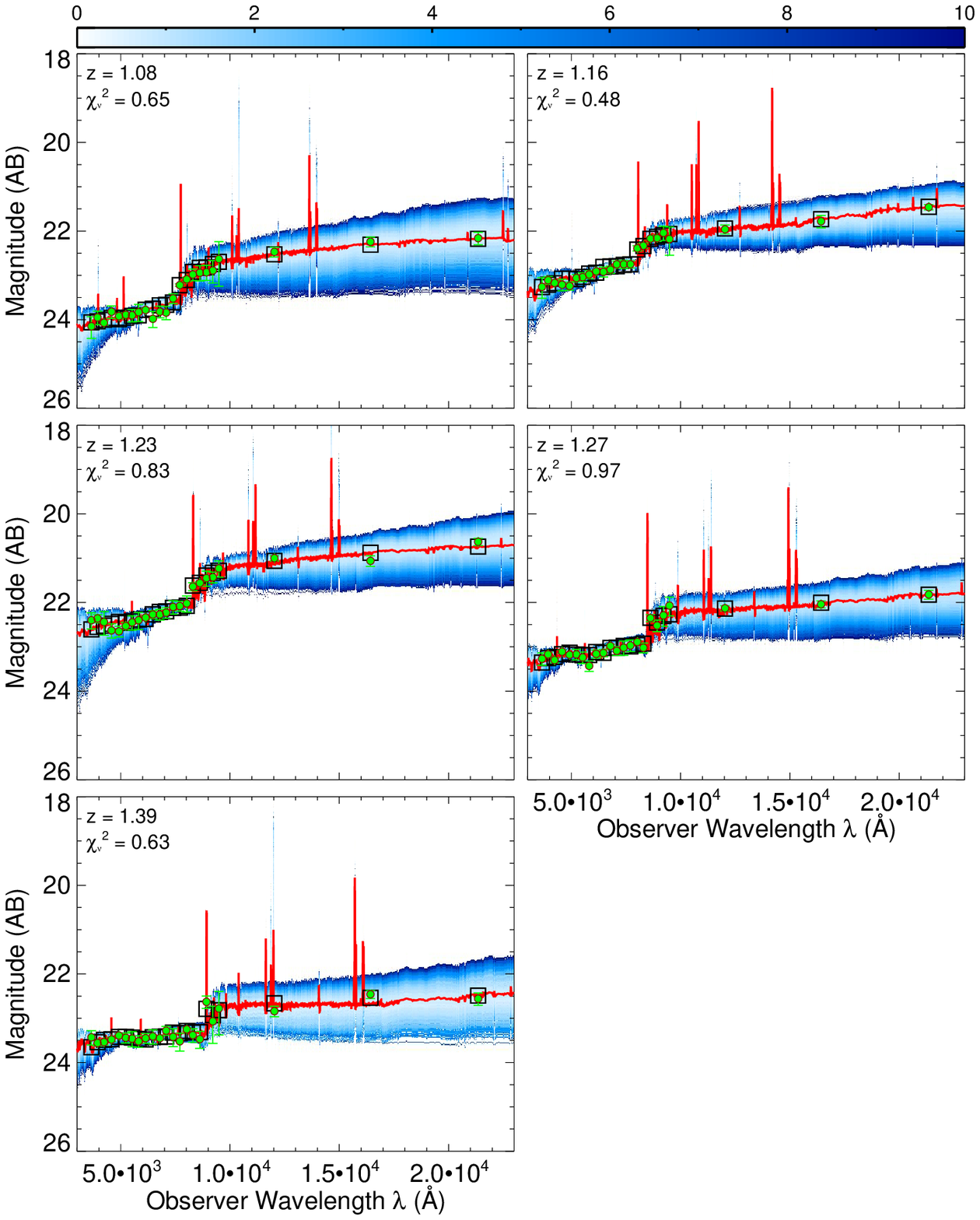}
\caption{Spectral Energy Distribution of a star-forming galaxy in the ALHAMBRA survey.
From left to right and up to bottom, the panel shows the SED of a galaxy picked out 
from the $BJG_7$,$BJG_8$,$BJG_9$,$BJG_{10}$, and $BJG_{11}$ samples, respectively.
The filled green dots show the ALHAMBRA photometric data.
The red line shows the model that minimizes the $\chi^2$ found by {\tt iSEDfit}, 
the minimum reduced $\chi^2$ is indicated in the upper left corner,
beside the BPZ photometric redshift.
The black squares mark the ALHAMBRA photometry of the best model.
The blue shading shows the Universe of models, generated by {\tt iSEDfit} (see Section \S\ref{s_isedfit}),
scaled by their reduced $\chi^2$. The color bar indicates the reduced $\chi^2$ scale.
}
\label{fig_isedfit2}
\end{figure*}

The figure \ref{fig_bias} shows the evolution of the bias factor as a function of redshift for the $BJGK$ samples (red squares).
The lines show the bias evolution for different halo masses \citep{sheth_mo_tormen01}, 
whose are indicated below of each line in the right side of the panel.
Approximately, the $BJGK$ samples follows the bias evolution of haloes of masses $\sim 10^{12.5} h^{-1} M_{\sun}$.
Finally, we calculate the halo mass using equation (8) of \cite{sheth_mo_tormen01} 
that relates the bias with the peak height, $\nu = \delta_{sc}(z)/\sigma(M,z)$, 
where $\delta_{sc}$ is the critical overdensity computed using the spherical collapse model.
Here we assume $\delta_{sc} = 1.69$. 
Then, the halo mass $M_h$ is obtained through $\sigma(M,z)$ evaluated at 
the redshifts of the $BJGK$ samples listed in table \ref{table_ppK}. 
In fig.~\ref{fig_ivan} the halo mass of the $BJGK$ samples as a function of redshift is shown.
Squares show the median values of the halo mass,
polygons show a conservative limit,
drawing the mass limits taking into account the $1\sigma$ deviation of the redshift and bias.
These results are presented in Table \ref{table_clustering}.

%__________________________________________________________________
\section{Discussion} \label{s_discussion}

%__________________________________________________________________

\subsection{Selection technique} \label{s_discussion_st}
The new two-color selection technique $U_2X_nK$ based on ALHAMBRA medium-bands,
allows us to extract eleven samples of star-forming galaxies at $z>$0.5
in narrower redshift ranges with respect to previous studies \citep{daddi04,steidel04,adelberger05}.
This selection method is based purely on the ALHAMBRA photometric data
and it does not depends on the models assumptions, methodology 
or templates used to determine the photometric redshifts 
or properties derived from SED fit.
We have validated this technique with the \cite{bc03} models as well as 
checked with other stellar population synthesis models giving consistent results.

The ALHAMBRA GOLD data is an $F814W$-selected catalog,
thus is less sensitive to detect the regions bluewards of the Balmer jump of galaxies at $z>$1. 
The $R_{12}$ band contributes only to $10\%$ of the final $F814W$ detection image, while  the $I_{13}$,  and $I_{14}$ bands contribute 18$\%$ each one.
Hence, the detection in the $F814W$ image becomes worse for galaxies at $z\ge$1 with a pronounced Balmer jump.

In order to avoid any contamination of low-redshift galaxies in our samples, we are inclined to use the bluest band available to sample the region bluewards of the Balmer jump.
Hence, we studied the detection level of the $U_1$ and $U_2$ that reaches $97\%$ and $99.7\%$, respectively.
Since the $U_2$ has a higher detection level with respect to $U_1$, we have tailored the selection technique using the $U_2$ band.

To select samples at $z<$0.5, we tried to use filters bluer than the $R_9$, whose central wavelength is lower than 613.5\AA.
Nevertheless, the theoretical evolution of the color $U_2X_{n<9}K$ of passive and star-forming galaxies,
based on the \cite{bc03} models, tend to occupy the same locus in the color-redshift ($U_2X_{n<9}K -\, z$) plane.
Hence, it does not allow to separate the star-forming from the passive galaxies as clear as for the $U_2X_nK$ selection with $n>9$ (see Fig. \ref{fig2} and Fig. \ref{fig3}).

The number density of each $BJG$ sample decreases with redshift,
probably due to the nature of these objects or 
also related to a non-considered Malquist-bias selection effect in the $U_2$-band. %and the $K_s$-band.
Besides, we verify our selection method with the BPZ photometric redshifts, 
whose uncertainty increase with redshift,
and therefore the number densities might be underestimated according to $\delta_z(BPZ) = 0.014\, z(BPZ)$.

According to this colour technique (visual inspection of Fig.\ref{fig2}, and Fig.\ref{fig3}),
the redshift distributions should be narrower than in Fig. \ref{fig_bpz}.
The distributions can become wider if the errors of the photometric redshifts
are properly taken into account.
The mean formal BPZ error of the $BJG_1$ and $BJG_{11}$ samples are 0.02 and 0.03,
corresponding to 20$\%$ and 30$\%$ of the total expected width ($\leq$ 0.1), respectively.
We performed simulations considering Gaussian redshift distributions
filled randomly using the same amount of galaxies found in each sample,
the expected width (0.1) and the formal BPZ error $\delta_z(BPZ) = 0.014(1+z(BPZ))$.
By perturbing the photometric redshift with its corresponding error,
and choosing randomly a positive or negative variance
the width of the distributions increase by a factor of 1.5 for the first $BJG_1$ sample,
and up to 2 times this value for the $BJG_{11}$ sample, with increasing redshift.
Clearly, higher photometric redshift errors increase the distribution width.
In the AHAMBRA data, BPZ tends to decrease its precision at I < 24.5 AB (Molino et al. 2014).
The global redshift distribution shows a mean of <z>$=0.86$ at this magnitude limit.
Hence the distributions at $z\geq 0.9$ tend to broaden even more because the 
photometric redshift errors are bigger than $0.014(1+z)$.
Considering a distribution at $z\sim1$ with an original width of 0.1 and $\delta_z(BPZ) = 0.03(1+z(BPZ))$, 
the width increase up to three times.
Outlayers due to either photometric redshift or
the color selection technique can also contribute to the tails 
as well as broader the photometric redshift distributions.

We have also investigated the selection method creating an $X_n$ band composed of two, 
three and four consecutive ALHAMBRA bands.
Indeed, it increases the statistics, accuracy, and width of the redshift distribution of each sample. Nevertheless, 
in order to fully exploit the capabilities of a multi-medium band survey in selecting galaxies in small redshift ranges,
we choose to use an unique band in our selection method.

\cite{viironen15} have shown that using the redshift probability distributions is possible to obtain a clean galaxy sample, 
which is by definition (intrinsically) not complete.
The fact of choosing certain probability threshold implies that
some galaxies will be not considered in the final sample. 
To overcome the purity versus completeness issue, 
they suggest to use the information of the whole zPDF to select each sample.
In this work we aim to tailor and use a two-color criteria that considers all the observational data available,
selecting a \emph{complete} sample, but at the same time is probably more \emph{contaminated} than 
the samples selected via the redshift probability distributions.
We have used the zBPZ-$BJG$ distributions to calculate this contamination,
by estimating how many objects have a BPZ redshift lower/higher than 3-$\sigma$
of its corresponding redshift distribution. 
Thus, it encompasses from 1$\%$ to 15$\%$ of each sample depending on redshift.
Specifically, 1$\%$, 2$\%$, 2$\%$, 3$\%$, 4$\%$, 5$\%$, 6$\%$, 7$\%$, 8$\%$, 10$\%$, 12$\%$,
and 15$\%$ of the sample $BJG_1$, $BJG_2$, $BJG_3$, $BJG_4$, $BJG_5$,
$BJG_6$, $BJG_7$, $BJG_8$, $BJG_9$, $BJG_{10}$, and $BJG_{11}$, respectively.

\subsection{Progenitors and descendants}  \label{s_discussion_c}
We have obtained the masses of the haloes where our $BJGK$ samples reside (see Fig.\ref{fig_ivan}).
The galaxy samples of similar $K$-band absolute magnitude 
reside on haloes of $10^{12.5 \pm 0.2} M_{\sun}$.
In terms of progenitors and descendants for the median halo mass, 
the progenitors are the LBGs with $R < 25$ at $z\sim$3 \citep{adelberger05,hildebrandt07}, 
the $sBzK$ galaxies with $K\leq23$ at $z\sim$2 \citep{hayashi07,hartley08,mccracken10,lin12} 
and the descendants in the local Universe are ``elliptical'' galaxies
with luminosities around 2 $L_*$ (see Fig.\ref{fig_bias}).
For haloes around the upper limit $10^{12.7} M_{\sun}$,
we find as progenitors at $z\sim$3 the LBGs \citep{hildebrandt07} with $R\leq 24$,
while at $z\sim$2 the sBzK galaxies with $K \leq 23$ \citep{hartley08,lin12} and
the descendants are elliptical galaxies with luminosities around 3 $L_*$.
For the haloes around the lower limit $10^{12.3 }  M_{\sun}$, 
the progenitors are the LBGs at $z\sim$4 \citep{kashikawa06,ouchi05} with I or z$\leq 27$,
the LAE galaxies at $z\sim$3 \citep{gawiser07},
the sBzK galaxies \citep{hayashi07} with $K<23$ at $z\sim$2,
the descendants are ``elliptical'' with luminosities around 1 $L_*$.
The {\it increment} of the median halo mass of $\sim$ 0.4~dex, between $z=1.0$ and $z=0.5$,
follows the observed mass increment in numerical simulations \citep[Millennium,][]{fakhouri10}.
It suggests that we are tracing the evolution of the haloes of masses around 
$\sim 10^{12.5}$ from $z=1.0$ to $z=0.5$,
and hence the evolution of the physical properties as it is shown in Fig. \ref{fig_ppevol}.
So far, via clustering measurements, we have argued that our five sets of galaxies $BJGK_{n<6}$ 
represent a coeval and homogeneous population of star-forming galaxies.

The bias and hence halo mass of the $BJGK$ samples are on average higher
than the one's found in \cite{hurtado15} for star-forming galaxies.
For all redshift ranges studied, our correlation lengths are ~$20\%$ higher 
with respect to their work 
suggesting that we are selecting more massive galaxies.
It might be due to the different galaxy selection methods as well as
the absolute magnitude thresholds used to define complete samples.
Nevertheless, the main difference relies on the method and
the scale that we have chosen to calculate the bias, and hence the halo mass.
In \cite{hurtado15}, they calculate the bias for the scale range $1.0 <r<10~$Mpc,
while in this work we evaluate the bias in the scale 8~$h^{-1}$Mpc in order to compare with previous works that also studied the progenitors and descendants of star-forming galaxies.
By taking this approach, we obtain halo masses 0.7~dex higher than in \citep{hurtado15}.

\subsection{Physical properties} \label{s_discussion_pp}
We perform an accurate estimation of the physical properties derived from SED fitting 
(e.g. absolute magnitude, stellar mass, star formation rate, etc.) 
and characterize each sample as a whole with the median of each properties.
Figure \ref{fig_ppevol} shows the median values of the probability distribution of stellar mass, 
absolute magnitude, star formation rate for the $BJGK_{n<6}$ (red squares)
samples as a function of redshift.
In the stellar mass panel, 
the dashed line shows the evolution of the stellar mass for galaxies 
with present-day stellar masses of $log(M_*)\approx 10.7$, as our Milky way galaxy,
taken from the analysis of the CANDELS survey by \cite{vandokkum13}.
There is a good agreement between the observed mass growth determined by \cite{vandokkum13},
which slightly increase by a factor of 0.1~dex from $z=1$ to $z=0.5$ 
and the flat behavior found in this work.
In the star formation rate panel, 
the dashed line shows the evolution of the implied star formation rate 
due to the evolution of the stellar mass of galaxies 
with present-day stellar masses of $log(M_*)\approx 10.7$,
as our Milky way, also taken from \cite{vandokkum13}.
Our results falls slightly below suggesting that $SFR$ is sufficient to 
account for the increment in stellar mass between $z=1$ and $z=0.5$ 
as well as major mergers play a minor role in the mentioned redshift range.
%SFR && sSFR
Yet, the evolution of the $SFR$ and $sSFR$ resembles the expected ``main-sequence'' behavior
of star-forming galaxies \citep{rodighiero11}.
%Luminosity
The $BJGK$ evolution, from $z\sim$1 to z=0, of the $B$-band absolute luminosity agrees with the 
evolution measured by \cite{zcosmos}. 
They found that the contribution of disks to the total $B$-band luminosity decreases by 30$\%$ 
from $z\sim$1 to z=0, while the $BJGs$ decrease their median $B$-band luminosity by a factor of 0.5~dex. 
The distribution of galaxy ages obtained by {\tt iSEDfit} has a large dispersion,
even so, its increment between two epochs corresponds to the Universe age increment indicated by the redshift.
According to the models of \cite{lagos14}, galaxies of properties similar to the $BJG$s host most of the neutral gas at $0.5<z<1.5$. 
Hence the $BJG$ samples contain tentative targets to sample the neutral gas with sub-millimeter surveys.
%__________________________________________________________________

\section{Conclusions}\label{s_conclusions}
%__________________________________________________________________

We have selected eleven samples of star-forming galaxies by using a two-color technique 
based on the \cite{bc03} models convolved with the ALHAMBRA filters.
Using clustering arguments, we have confirmed that 
five out of the eleven sets of galaxies, i.e. the $BJGK_{n<6}$,
represent a coeval and homogeneous population of star-forming galaxies.
The properties derived from SED fitting, such as stellar mass, star formation rate, age, absolute luminosity,
of each $BJGK_{n<6}$ sample are characterized as a
whole allowing us to study their putative evolution as a function of redshift.
The main results can be summarized as follows:

\begin{itemize}
 \item We tailor a two-color selection technique, 
 based on the \cite{bc03} models and the Balmer jump that select star-forming galaxies in the redshift range $0.5<z<1.5$.
 We select eleven samples composed of Balmer jump Galaxies, dubbed $BJG$. 
 The amount of photometric-redshift outliers in these color-selected samples increases with redshift ranging from 1$\%$ to 15$\%$.
 \item We create a sub-sample of $BJG$, dubbed $BJGK_n$ with $n<6$, 
 which considers only the $BJG_{n<6}$ galaxies brighter than $K_{abs}\sim-21.2$.
 The stellar mass of the $BJGK$ samples nearly does not change with redshift,
 suggesting that major mergers played a minor role on the evolution of the $BJGK$ galaxies.
 The SFR evolution accounts for the small variations of the stellar mass, 
 from $z \sim 1$ to $z \sim 0.5$,
 suggesting that star formation and minor mergers are the main channels of mass assembly.
Although the distribution of galaxy ages obtained by {\tt iSEDfit} has a large dispersion, the evolution of the galaxy age agrees with the evolution of the Universe age.
  
 \item The $BJGK_{n<6}$ samples reside on haloes of $10^{12.5 \pm 0.2} M_{\sun}$,
which progenitors are the LBGs with $R \leq 24$ at $z\sim$3 \citep{hildebrandt07}, 
the $sBzK$ galaxies \citep{lin12,hartley08,mccracken10} with $K\leq23$ at $z\sim$2
and descendants are elliptical galaxies with luminosities around 1-3 $L_*$ 
(for details see Fig.\ref{fig_bias}).

 \item The similar increment of the median halo mass between $z=1.0$ and $z=0.5$
of our observational results and numerical simulations \citep[Millennium,][]{fakhouri10}
suggest that we are tracing the evolution of haloes of masses around $\sim 10^{12.5}$ from $z=1.0$ to $z=0.5$, and hence the putative evolution of the physical properties 
of the galaxies hosted by these haloes (see Fig. \ref{fig_ppevol}).
 \end{itemize}

The homogeneous coverage of the ALHAMBRA optical bands from $R_9$ to $z_{20}$,
allows us to trace the evolution of the baryonic processes occurring on star-forming galaxies,
from $z\sim 1.0$ to $z\sim 0.5$, which reside on haloes of masses $\sim 10^{12.5} h^{-1} M_{\sun}$.
It corresponds roughly to a stellar mass upper (lower) limit of $\sim 10^{11} M_{\sun} (10^{10} M_{\sun})$.
Deeper ALHAMBRA data as well as {\it near-infrared selected catalogs}, 
would allow us to study the evolution of the physical properties on haloes of masses lower
than $10^{12.5} M_{\sun}$ from $z\sim 0.5$ to $z\sim 1.5$. 

\hskip 1.5in

\begin{acknowledgements}
The work presented in this paper is based on observations 
taken at the Centro Astron\'omico Hispano Alem\'an (CAHA) at
Calar Alto, operated jointly by the Max-Planck Institut f\"{u}r 
Astronomie and the Instituto de Astrof\'isica de Andaluc\'ia (IAA–CSIC).
PT and AMA acknowledge support from FONDECYT 3140542 and FONDECYT 3160776, respectively.
LI, AMA, PT, SG, NP acknowledge support from Basal-CATA PFB-06/2007.
EJA acknowledges support from the Spanish Ministry for Economy and Competitiveness and FEDER funds through grant AYA2013-40611-P.
AFS, VJM and PAM acknowledge partial financial support from the  Spanish Ministry for Economy and Competitiveness and
FEDER funds through grant AYA2013-48623-C2-2, and from Generalitat Valenciana through project PrometeoII 2014/060.
\end{acknowledgements}

\bibliographystyle{aa} % style aa.bst
%   \bibliography{aa_1109.bib} % your references Yourfile.bib
\bibliography{references}
\end{document}